\documentclass{PoS}

\title{Extracting resonance parameters from lattice data}

\ShortTitle{Extracting resonance parameters from lattice data}

\makeatletter
\renewcommand\speaker[1]{\if@speaker\global\@dblspeaktrue\fi
                        \global\@speakertrue
                        \global\setbox\@firstaubox
                        \hbox{{\let\thanks\@gobble
                                \let\footnote\@gobble\small 
                                \rm #1}}%
                        #1\thanks{Speakers.}\
                        }%
\makeatother

\author{\speaker{Pietro Giudice}, Darran McManus$^*$ and Mike Peardon\\
  School of Mathematics, Trinity College, Dublin 2, Ireland\\
  E-mail: \email{giudice@maths.tcd.ie},
  \email{dmcmanu@tcd.ie},
  \email{mjp@maths.tcd.ie}
}

\abstract{
Monte Carlo simulations of the 4d O(4) model in the broken
phase are performed to determine the parameters of a resonance.
The standard method for extracting them on the lattice is
through L\"uscher's formula; recently a new method, based on the
probability distribution concept, has been proposed.
We study the application of these methods and compare them
with Monte Carlo data.
}

\FullConference{The XXVIII International Symposium on Lattice Field Theory, 
Lattice2010\\
		June 14-19, 2010\\
		Villasimius, Italy}

\usepackage{amsfonts}
\usepackage{setspace}
\usepackage{url}
\usepackage{graphicx}
\usepackage{psfrag}
\newcommand {\onehalf}{\frac{1}{2}}

\begin{document}

\section{Introduction}

In Quantum Chromodynamics, the current theory of the strong force, 
many states do not appear directly in scattering 
experiments, but only indirectly in the behaviour of the scattering cross 
sections of observable particles. This is because these states, known as 
resonances, are unstable and decay in a very short time relative to the 
scattering experiment. Extracting the decay widths and masses 
of these states is thus an important theoretical challenge.

Lattice Field Theory provides a possible way of extracting these resonance 
parameters in a non-perturbative fashion. The typical method for extracting 
particle masses in lattice field theory is to study the decay of a 
correlator with the same quantum numbers as the particle in question. 
The large-time behaviour of the correlator is then 
\begin{equation}
\lim_{t\rightarrow\infty} C(t) = Z e^{-mt} \ ,
\end{equation}
where $m$ is the mass of the lightest particle with the chosen quantum
 numbers, which can be extracted by fitting the correlator.
This method however will not work for resonances. By virtue of being 
unstable, resonances are above the multiparticle threshold in 
their channel and never dominate the behaviour of the correlator in a 
simple manner. Fundamentally resonances are not energy eigenstates of 
the Hamiltonian, but rather poles of the S-matrix and so are a truly 
dynamical phenomena.

Given their relation to dynamical scattering processes, resonance 
parameters can be found using the scattering phase shift $\delta(p)$. 
The difficulty lies in obtaining information about $\delta(p)$ on 
a Euclidean lattice. In Ref~\cite{L\"uscher:1991a} it was discovered that 
there is a connection between the behaviour of the two particle 
energy spectrum in finite volume and the scattering phase shift 
$\delta(p)$. Hence provided one can accurately determine the energy 
spectrum it should be possible to obtain $\delta(p)$ and through it 
the resonance parameters.
Recently another method has been proposed in Ref~\cite{Bernard:2008ax}. 
This method takes the intuition gained from L\"uscher's method to 
construct a probability distribution
which measures the relative frequency of energy levels in the 
interacting case (resonance present) and the noninteracting case 
(no resonance); we will refer to this method as the \emph{histogram method}. 
It can be shown that the parameters of this 
probability distribution are fundamentally related to the parameters 
of the resonance. Furthermore, the method provides a visual tool. 
The resonance should manifest itself as a peak in the distribution.

Our first discussion on this topic, with an historical introduction and 
a basic theoretical background, can be found in Ref~\cite{GiudicePeardon:2009}.
In this work we aim to compare and contrast these two methods. 
In particular we analyse how accurately resonance parameters can be 
extracted using both methods and also the ambiguity in applying 
the two methods.
A first attempt to test the histogram method on a simple 
one dimensional model can be found in Ref~\cite{Morningstar:2008mc}.

\section{Theoretical background}

\subsection{Two particles in a box}

In the continuum, two identical non-interacting bosons of
mass $m_\pi$ characterised by a relative momentum $\vec{p}$, 
in a box of
volume $V=\prod_{i=1}^3 L_i$, have a total energy $E$ given by
\begin{equation}
E=2 \sqrt{m_\pi^2+\vec{p}^2} \ ;
\label{eqEcont}
\end{equation}
where due to the finite volume, the momenta $p_i$ are given by
$p_i=\frac{2 \pi}{L_i} n_i$ with $n_i \in Z$.
On the lattice the space-time discretization can have a strong effect
(see Figure~\ref{spectr_cont_latt})
in particular when the volume is small (large momentum) and $m_\pi$ is big.
The correct expression for the simplest discretisation of the free scalar field
is 
\begin{equation}
E=4 \sinh^{-1}\left[{\onehalf \sqrt{m_\pi^2+\tilde{p}^2}}\right] \ ,
\label{eqElatt}
\end{equation}
where $\tilde{p}_i=2 \sin{\frac{\pi}{L_i}n_i}$. It is also valuable to use this
expression to describe the energy of interacting particles as we will show. 
\FIGURE{
  \psfrag{LL}{$L$}
  \psfrag{EE}{$E_n$}
  \psfrag{MM}{$m_\pi=0.46$}
  \includegraphics[width=0.36\textwidth,angle=-90]
{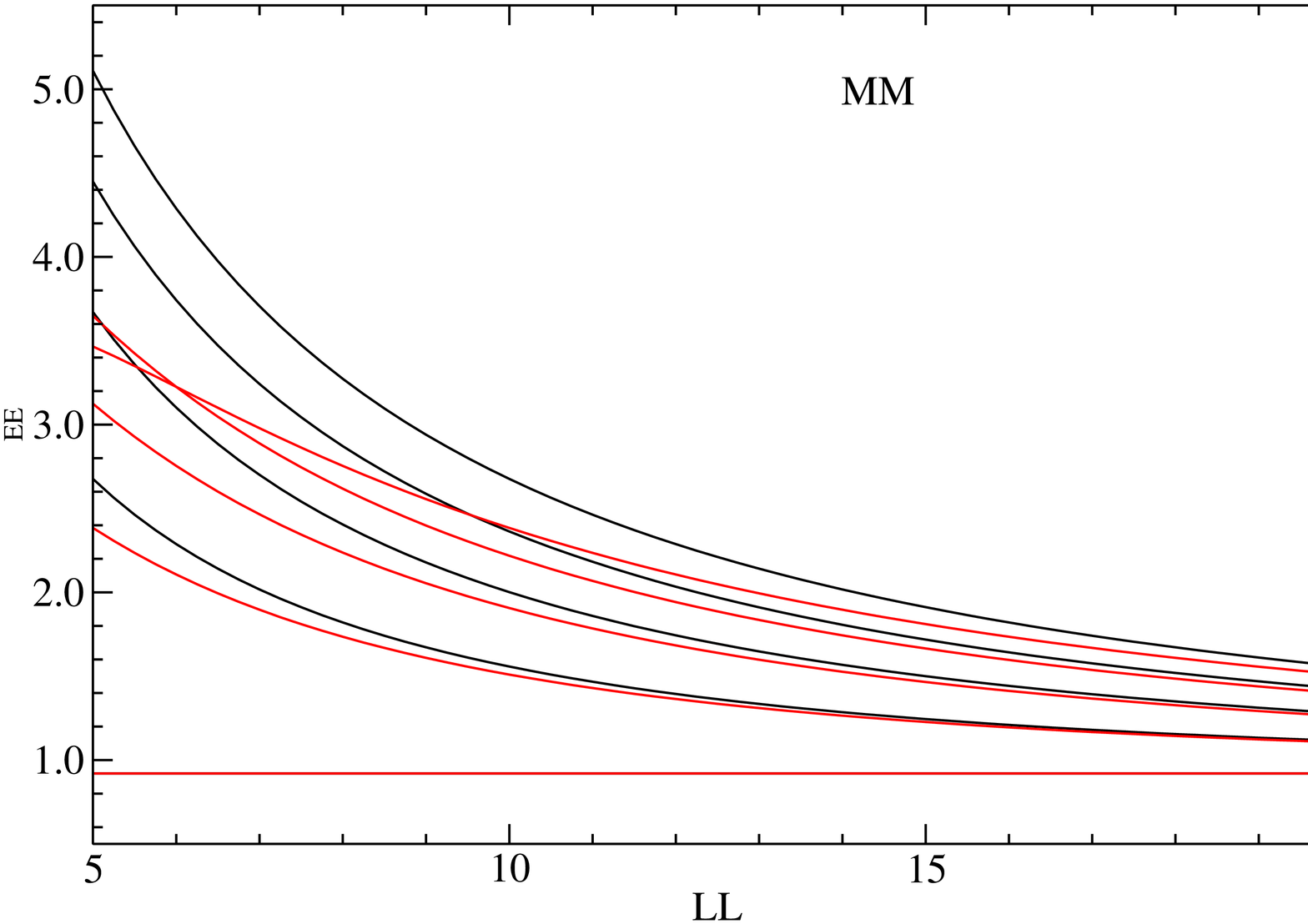}
\caption{The total energy $E$ for four different levels in the continuum 
(black lines) and in the lattice (red lines) case versus $L$.}
\label{spectr_cont_latt}
}
In a general case such as QCD, where an expression like 
Eq.~\protect\ref{eqElatt} is not available, we need to determine 
the non-zero-momentum single-particle energy levels 
numerically and then, to determine the 
two-particle energy spectrum, simply multiply the results by a factor of two.

We will focus on the case of a cubic lattice, 
characterised by a single side length $L$; moreover,
in a cubic box if $n^2=\sum_{i=1}^3 n_i^2$ is fixed, degenerate
energy levels for different values of $n_i$ can appear.

In Figure~\ref{spectr_cont_latt} we show a plot of the two formulas where
it is evident that for small volume ($L \lesssim 15$) and a mass $m_\pi=0.46$
the two spectra are very different; therefore we cannot use the continuum
formula to describe our Monte Carlo results.

\subsection{Avoided level crossing}
\begin{figure}[ht]
  \footnotesize
  \psfrag{L}{L}
  \psfrag{E}{E}
  \psfrag{PHI}{$m_\phi$}
  \hspace{-2mm}
  \includegraphics[width=0.36\textwidth,angle=-90]
{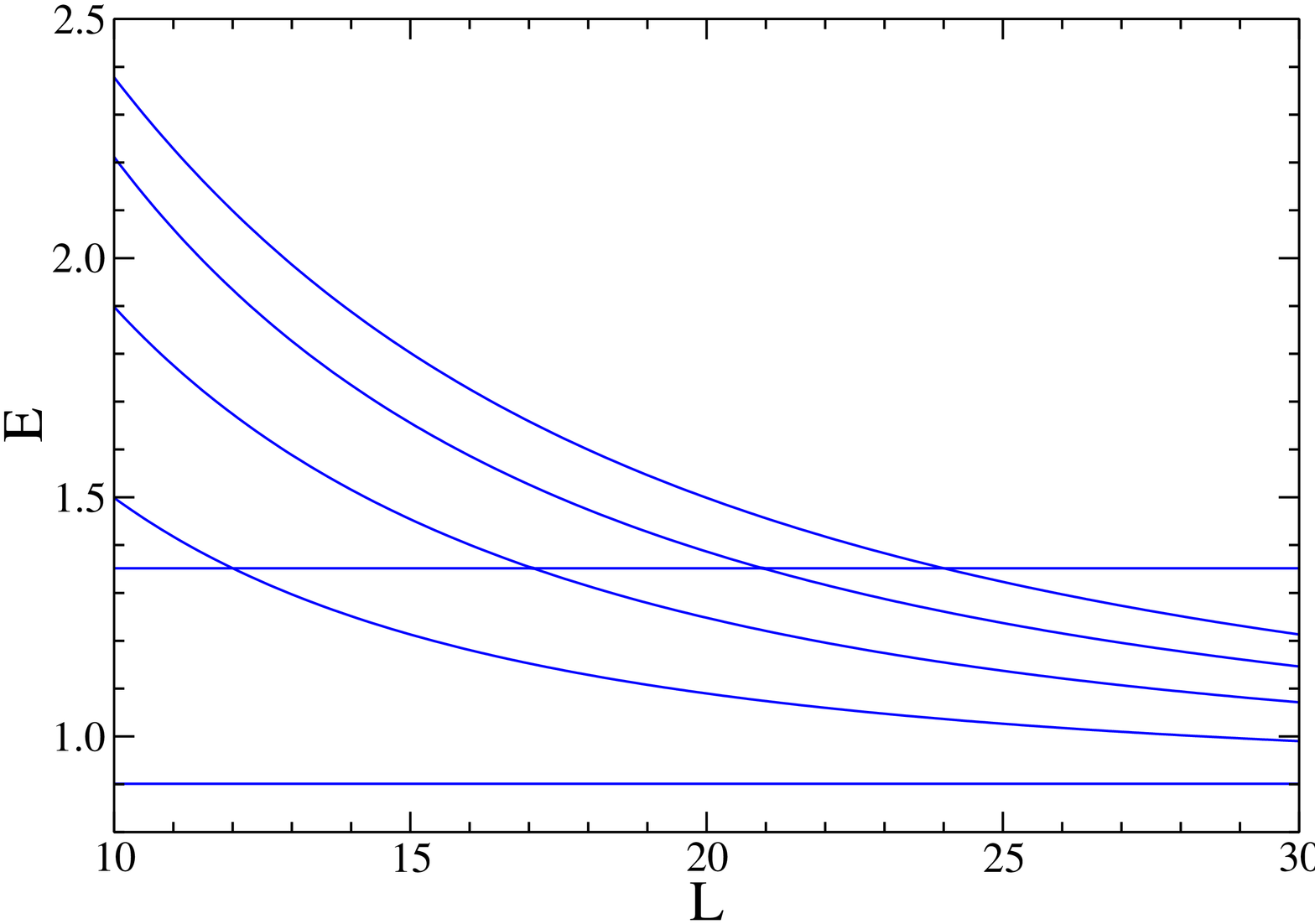}
\hspace{2mm}
  \includegraphics[width=0.36\textwidth,angle=-90]
{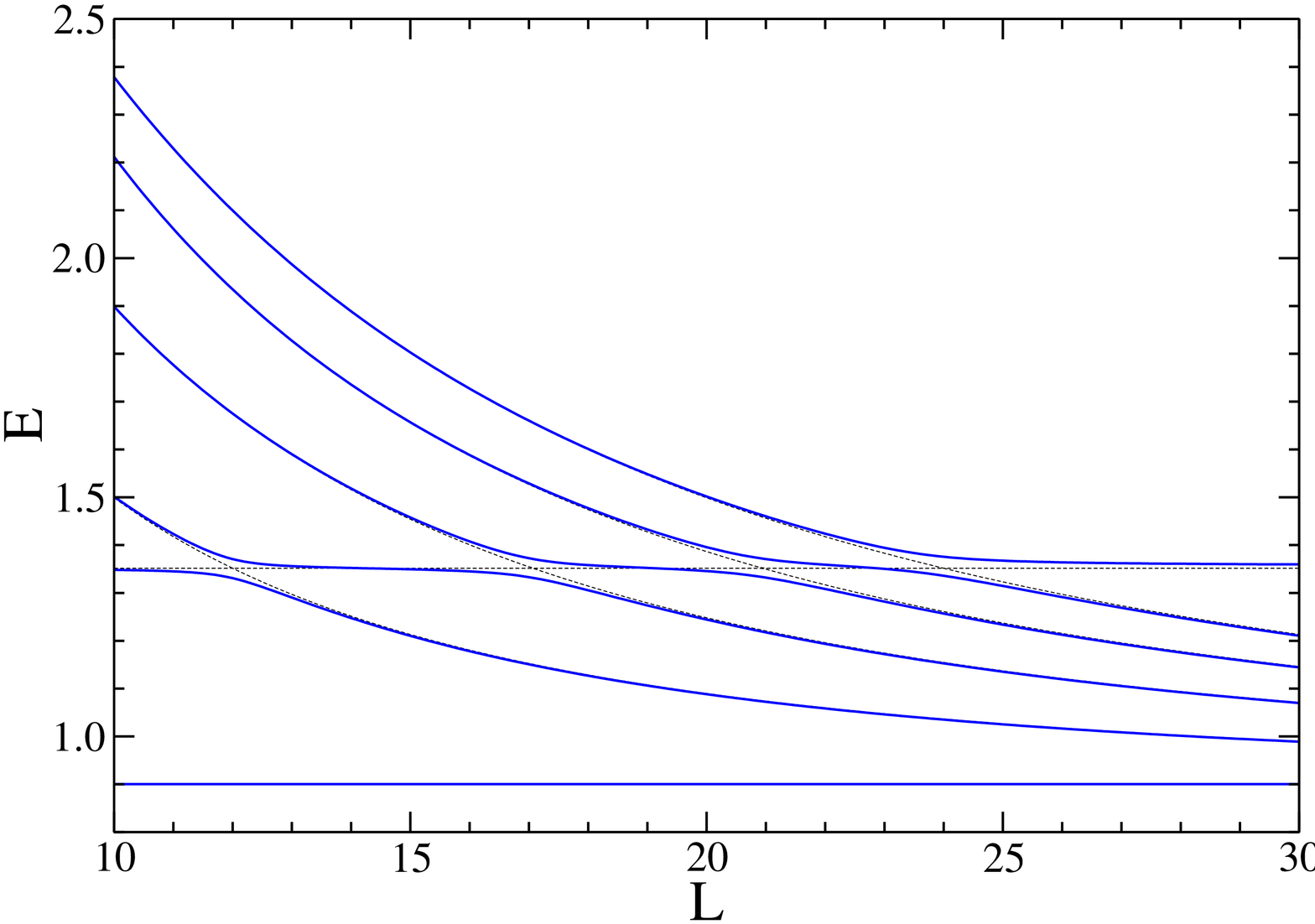}
\caption{(Left) The spectrum of a system of two non-interacting particles
of mass $m_\pi=0.4544$ worked out using Eq.~\protect\ref{eqElatt}; the
horizontal line describes the particle $\sigma$ at rest with mass
$m_\sigma=1.3517$.
With these parameters the intersection between $\sigma$ and the two-particle
level $n^2=1$, i.e. $(1,0,0)$, is set at $L=12$. (Right) Avoided level
crossings where on the (Left) there were intersections between $\sigma$
and $2\pi$.}
\label{spectrum_tot}
\end{figure}

Let us introduce another particle $\sigma$ in the box (at the moment,
not interacting)
with mass $m_\sigma$; we are interested in studying the 
\emph{elastic} scattering between the $\pi$ particles therefore we 
impose the constraint $2 m_\pi < m_\sigma < 4 m_\pi$.
In Figure~\ref{spectrum_tot} (Left)
the $\sigma$ energy level is the horizontal line that
intersects the two-particle levels at various system sizes $L$.

In Minkowski space if we introduce a three point interaction $\sigma\pi\pi$ 
between the fields, the $\sigma$ can be unstable and decay into two 
$\pi$-particles.  In Euclidean space and in a finite volume the scenario is 
different; because of the interaction, the energy eigenstates are a mixture of
$\sigma$ and $2\pi$ Fock-states and they are all stable.
The mixing is manifested in avoided level crossings (ALCs)
in the energy levels as shown in Figure~\ref{spectrum_tot} (Right).

\subsection{L\"uscher's method}

The best known method for analysing resonances was proposed by 
L\"uscher (Ref~\cite{L\"uscher:1991b}). This method involves using information
on the scattering phase shift. Since the scattering phase 
shift contains information on resonance parameters this provides a way to 
extract the resonance mass and width. The scattering phase shift itself can be 
obtained using the relationship found in (Ref~\cite{L\"uscher:1991a}). In 
essence, this relationship provides a mapping between the values of the 
two-particle energy spectrum in finite volume and the scattering phase shift 
in infinite volume.

The relationship is proven first in non-relativistic quantum mechanics
It then holds in quantum field theory as the relativistic case can be cast
in a non-relativistic form, with the Bethe-Salpeter Kernel playing the r\^ole
of the potential. This is achieved by means of an effective Sch\"odinger
equation, first constructed in (Ref~\cite{L\"uscher:1986}).
The precise relationship is\footnote{In truth the relationship is more 
general than this, the formula quoted here is for the spin-$0$ channel, 
which is the only one relevant here.}
\begin{eqnarray}
\delta(p) = -\phi(\kappa) + \pi n \ , 
\label{lusherformu}
\\
\tan(\phi(\kappa)) = \left( \frac{\pi^{3/2}\kappa}{Z_{00}(1;\kappa^{2})} 
\right), \quad \kappa = \frac{pL}{2\pi} \ ,
\label{tanlusherformu}
\end{eqnarray}
where $p$ is the relative momentum between the two pions. 
$\mathcal{Z}_{js}(1;q^{2})$ is a generalised Zeta function, given by
\begin{equation}
\mathcal{Z}_{js}(1;q^{2}) = \sum_{\underline{n} \in \mathbb{Z}^{3}} 
\frac{r^{j}Y_{js}(\theta,\phi)}{(\underline{n}^{2} - q^{2})^{s}} \ ,
\end{equation}
where $Y_{js}(\theta,\phi)$ are the spherical harmonics. 
Eq.~\ref{lusherformu} is known as L\"uscher's formula.

To obtain resonance parameters using this relationship one proceeds as follows:
\begin{enumerate}
\item Through Monte Carlo simulations, obtain the two particle energy spectrum
$E_n(L)$ at different volumes;
\item Through dispersion relations obtain a momentum from the energy 
spectrum, $p_n(L)$;
\item Through Eq.~\ref{lusherformu} $p_n(L)$ gives a value for $\delta(p_n(L))$;
\item By repeating this procedure for several energy levels and volumes, 
one obtains a profile of $\delta(p)$;
\item This profile of $\delta(p)$ is then fitted against the Breit-Wigner 
form for the scattering phase shift in the vicinity of a resonance:
\begin{eqnarray}
\delta(p) \approx \tan^{-1} \left( \frac{4p^{2} + 4M_{\pi}^{2} - M_{\sigma}^{2}}
{M_{\sigma}\Gamma_{\sigma}} \right) \ .
\end{eqnarray}
This fit should give the resonance mass $M_{\sigma}$ and width $\Gamma_{\sigma}$.
\end{enumerate}
The work in this paper applies this method to the the $O(4)$ model in the 
broken phase and also compares its performance against a more recent
proposal.

\subsection{Histogram method}
\label{subsec_propmeth}

An alternative method to determine the parameters of a resonance is
based on a different way to \emph{analyze} the finite volume energy
spectrum (Ref~\cite{Bernard:2008ax}). The basic idea is to construct the 
probability distribution $W(E)$ according to the prescriptions:
\begin{enumerate}
\item Measure the two-particle spectrum $E_n(L)$ for different values of $L$
and for $n=1, \cdots, N$ levels;
\item Interpolate the data with fixed $n$ in order to have a continuum
function $E_n(L)$ in an entire range $L \in [L_{0},L_{M}]$;
\item Slice the interval $[L_{0},L_{M}]$ into equal parts with length
$\Delta{L} = (L_M-L_0)/M$;
\item Determine $E_n(L_i)$ for each $L_i$ ($i=0, \cdots, M$);
\item Choose a suitable energy interval $[E_{min},E_{max}]$ and introduce an
equal-size energy bin with length $\Delta{E}$;
\item Count how many eigenvalues $E_n(L_i)$ are contained in each bin;
\item Normalize this distribution in the interval $[E_{min},E_{max}]$.
\end{enumerate}

Actually, the distribution considered in Ref~\cite{Bernard:2008ax} is $W(p)$
but this does not have an important effect on our analysis; as a matter of fact,
the relation between them is:
\begin{equation}
W(p)=W(E) \left( \frac{\partial{E}}{\partial{p}} \right) \ ,
\label{wewp}
\end{equation}
where the correct dispersion relation we should use is Eq.~\ref{eqElatt};
the multiplicative term will not modify the Breit-Wigner shape \emph{near}
the resonance.

It is possible to show that the probability distribution $W(p)$ is given
by $W(p)=c \sum_{n=1}^N \left[ p^\prime_n(L) \right]^{-1}$ and differentiating
the L\"uscher formula with respect to $L$, it turns out
($c$ is a normalization constant):
\begin{equation}
W(p)=\frac{c}{p} \sum_{n=1}^N \left[L_n(p)+ \frac{2 \pi \delta^\prime(p)}
{\phi^\prime(q_n(p))} \right] \ .
\label{probdistr}
\end{equation}
The authors of Ref~\cite{Bernard:2008ax} introduce $W_0(p)$, which is determined
by  Eq.~\ref{probdistr} with $\delta(p)=0$ and $L_n(p)$ corresponding to the
free energy levels: they show that in order to subtract the
background (free $\pi$ particles) it is convenient to consider the
subtracted probability distribution $\tilde{W}(p)=W(p)-W_0(p)$.

Using convenient approximations in the L\"uscher formula and in the limit
of infinite number of energy levels (infinite volume) it turns out:
\begin{equation}
W(p)-W_0(p) \propto \frac{1}{p} \left( \frac{\delta(p)}{p} -
  \delta'(p) \right) \ .
\label{wpmwp0}
\end{equation}

This last quantity is determined by $\delta(p)$ alone and close to the
resonance, assuming a smooth dependence on
$p$ for the other quantities, it follows the Breit-Wigner shape of the
scattering cross section with the same parameters:

\begin{equation}
W(p)-W_0(p) \propto \frac{1}{[E(p)^2-M_\sigma^2]^2+M_\sigma^2 \ \Gamma^2} \ .
\label{bw}
\end{equation}

In Ref~\cite{Bernard:2008ax} this method is tested on \emph{synthetic} data
produced using the L\"uscher formula by experimentally measured phase shifts.
The main task of our work is to test this method on an effective field theory
where a resonance emerges, producing data by lattice simulations.

\section{The model}

The model we have used in our simulations is essentially the $O(4)$ model
in the broken phase. This model has previously been used to test L\"uscher's 
method (Ref~\cite{Gockeler:1994}). The Lagrangian is the following:
\begin{equation}
\mathcal{L}=\onehalf \partial \phi_i \partial \phi_i + \lambda
( \phi_i^2-\nu^2 )^2-m_\pi^2 \nu \phi_4 \ , \qquad \mbox{with i=1,2,3,4} \ .
\label{lagr0}
\end{equation}
The term proportional to $\phi_4$ is introduced to break explicitly the
symmetry and therefore to give mass to the three Goldstone bosons.
To understand the meaning of the terms and the parameters in the Lagrangian,
we first introduce the new fields $\sigma$ and $\rho_i$
(with the constraint $\rho_i \rho_i=1$):
\begin{equation}
\phi_i=(\nu+\sigma) \rho_i \ ,  \qquad \mbox{with $i=1,2,3,4$} \ ;
\label{newfileds}
\end{equation}
then, we expand the potential around
the classical minimum $\phi_i\phi_i =\nu^2$
(using also $\rho_i \partial \rho_i=0$):
\begin{eqnarray}
\mathcal{L} &=& \onehalf \nu^2 \partial \rho_i \partial \rho_i +
\onehalf \sigma^2 \partial \rho_i \partial \rho_i +
\onehalf \partial \sigma \partial \sigma +
\nu \sigma \partial \rho_i \partial \rho_i \nonumber \\
&+&
\lambda \sigma^4 +4\nu^2 \lambda \sigma^2+4 \nu \lambda \sigma^3
-m_\pi^2 \nu^2 \rho_4-m_\pi^2 \nu \sigma \rho_4 \ .
\label{lagr1}
\end{eqnarray}
The field $\sigma$ is clearly related to the massive field whereas the
four constrained fields $\rho_i$ are related to the three ``pions''. There is 
an easy way
to see this based on the treatment of the non-linear sigma 
model\footnote{See for example Ref~\cite{DeGrand:2006zz} Sec~2.4.3.}; 
we introduce the pions using an element of $SU(2)$:
$U=\exp{\left( \frac{i}{f}\pi_j \sigma_j \right)}$, where $\sigma_j$ are the
three Pauli matrices and $f$ is the pion decay constant. It turns out that
\begin{equation}
{\onehalf \mbox{Tr} \left( \partial_\mu U \partial_\mu U^\dagger \right)}
\stackrel{f \to \infty}{\Rightarrow}
{\frac{1}{f^2}\sum_{j=1}^3 \partial_\mu \pi^j \partial_\mu \pi^j} \ .
\label{rel1}
\end{equation}
On the other hand, we can introduce an element of $SU(2)$ by
$U=\rho_4+i \sigma_j \tilde\rho_j$, with $j=1,2,3$ and the constraint
$\rho_4^2+\tilde\rho_j \tilde\rho_j =1$; in this case we have:
\begin{equation}
{\onehalf \mbox{Tr} \left( \partial_\mu{U} \partial_\mu{U^\dagger} \right)} =
\partial \rho_4 \partial \rho_4+ \sum_{j=1}^3 \partial \tilde\rho_j
\partial \tilde\rho_j=
{\sum_{i=1}^4 \partial \rho_i  \partial \rho_i} \ .
\label{rel2}
\end{equation}
Comparing Eq.~\ref{rel1} and Eq.~\ref{rel2} it turns out ($f \to \infty$):
\begin{equation}
\sum_{i=1}^4 \partial \rho_i  \partial \rho_i \simeq
\frac{1}{f^2} \sum_{j=1}^3 \partial \pi_j \partial \pi_j  \ .
\label{relfin}
\end{equation}
Using the same previous approach and the relation $\mbox{Tr} ( U + U^\dagger )$
it is easy to show that
\begin{equation}
\rho_4= - \frac{1}{2 f^2} \pi_j \pi_j + const
\label{relrho4}
\end{equation}
We can rewrite the Lagrangian of Eq.~\ref{lagr1} using
Eq.s~\ref{relfin}-\ref{relrho4}
and introducing $\tilde\pi_j=\pi_j \frac{\nu}{f}$:
\begin{eqnarray}
\mathcal{L}&=&\onehalf \partial \tilde\pi_j \partial \tilde\pi_j +
\frac{1}{2 \nu^2} \sigma^2 \partial \tilde\pi_j \partial \tilde\pi_j +
\onehalf \partial \sigma \partial \sigma +
\frac{1}{\nu} \sigma \partial \tilde\pi_j \partial \tilde\pi_j
\nonumber \\
&+&
\lambda \sigma^4 +4\nu^2 \lambda \sigma^2+4 \nu \lambda \sigma^3
+\frac{1}{2} m_\pi^2 \tilde\pi_j \tilde\pi_j+
\frac{m_\pi^2}{2 \nu} \sigma \tilde\pi_j \tilde\pi_j \ .
\end{eqnarray}
In this expression, it is now evident that $\tilde\pi_j$ are the pions
with mass $m_\pi$ and $\sigma$ is a massive field
with mass $m_\sigma=2 \nu \sqrt{2 \lambda}$.
It is interesting to note there are two three-point interaction terms, both
inversely proportional to $\nu$.

\section{Monte Carlo simulation}

The theory described by the Lagrangian Eq.~\ref{lagr0} was simulated using
an overrelaxation algorithm for the first three fields followed by a
Metropolis update to guarantee the ergodicity and a Metropolis
algorithm for the field $\phi_4$.

\subsection{Single and two-particle spectrum}

In order to determine the single particle spectrum we first introduce the
partial Fourier transform (PFT) of the four fields $\phi_i$:
\begin{equation}
\tilde{\phi_i}(\vec{n},t)=\frac{1}{V}\sum_x \phi_i(\vec{x},t) e^{-i \vec{x} \vec{p}} \ , \qquad
p_i=\frac{2 \pi}{L_i} n_i \ , \qquad n_i=0, \cdots ,L_i-1 \ .
\end{equation}
The single particle mass is extracted from the zero momentum correlation
function ($\vec{n}=\vec{0}$):
\begin{equation}
C_i(t)= \langle \tilde{\phi_i}(\vec{n},t) \tilde{\phi_i}(-\vec{n},0) \rangle \ ,
\end{equation}
in particular with $i=1,2,3$ we can determine the mass of the $\pi_i$
particles; with $i=4$ we extract the mass of the $\sigma$ particle.
Because of the different way we update the four fields, it turns out that
the mass $m_\pi$ is determined with a higher precision then $m_\sigma$;
actually, this is not a real problem because we are mainly interested
to a good determination of the ``pion'' mass.

The two-particle spectrum is measured by introducing operators with
zero total momentum and zero isospin:
\begin{equation}
O_{\vec{n}}(t)= \sum_{i=1}^3 \tilde{\phi_i}(\vec{n},t)
\tilde{\phi_i}(-\vec{n},t) \ ;
\end{equation}
we take in account $N-1$ different operators, corresponding to
$n^2=0,1,\cdots,N-1$.
A $N$-th operator, that clearly has the correct quantum number, is the PFT
of the field $\sigma$ (actually $\phi_4$) with $\vec{p}=0$.
To determine the energy levels we use a method, introduced
in Ref~\cite{Luscher:1990ck}, based on a generalized eigenvalue problem applied
to the correlation matrix function $C_{ij}(t)= \langle O_i O_j\rangle$,
i.e. a matrix whose elements are all possible correlators between the $N$
operators.

\subsection{Numerical results}

\begin{figure}[htb]
\hspace{-2mm}
  \includegraphics[width=0.36\textwidth,angle=-90]{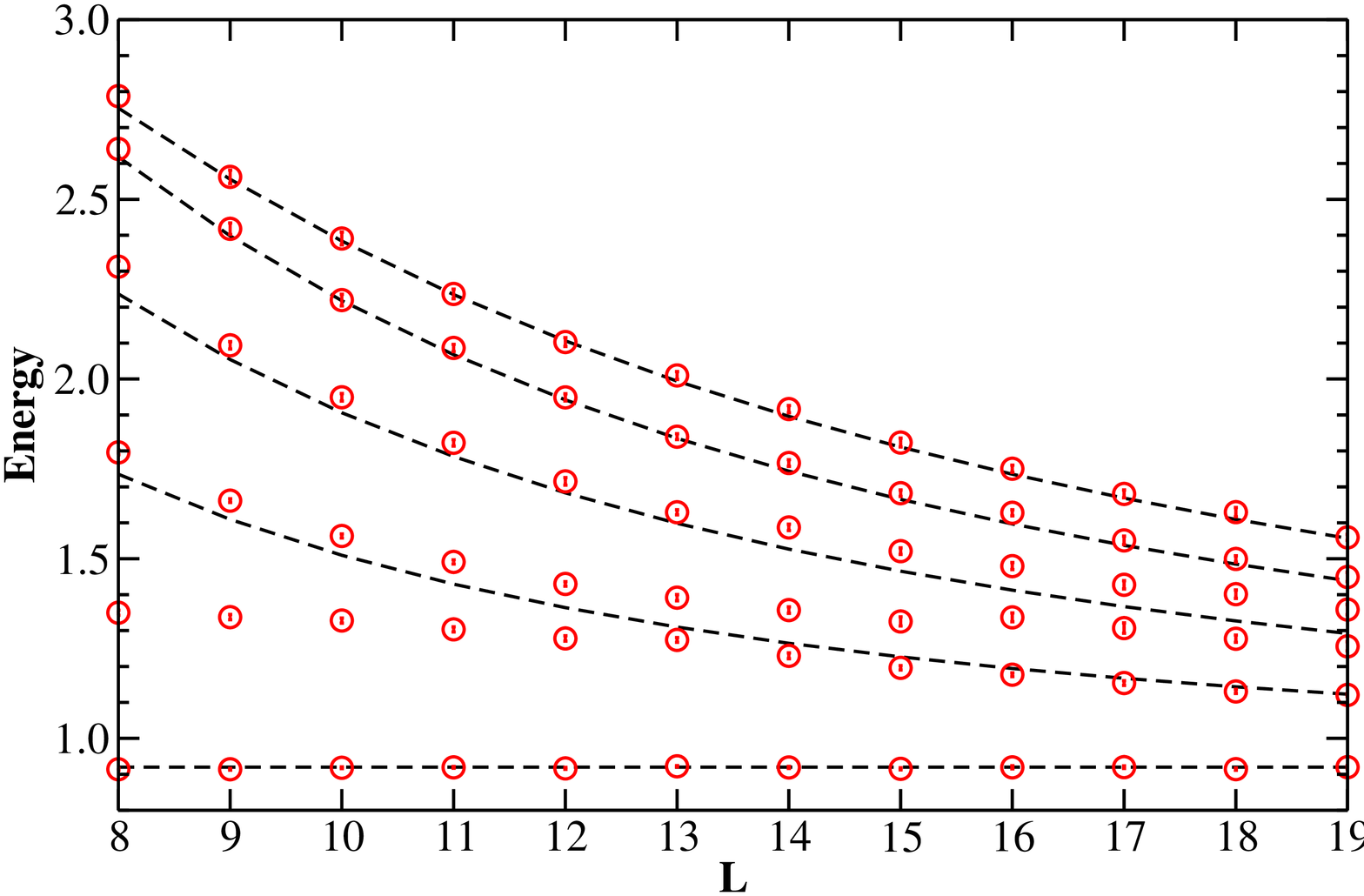}
\hspace{2mm}
  \includegraphics[width=0.36\textwidth,angle=-90]{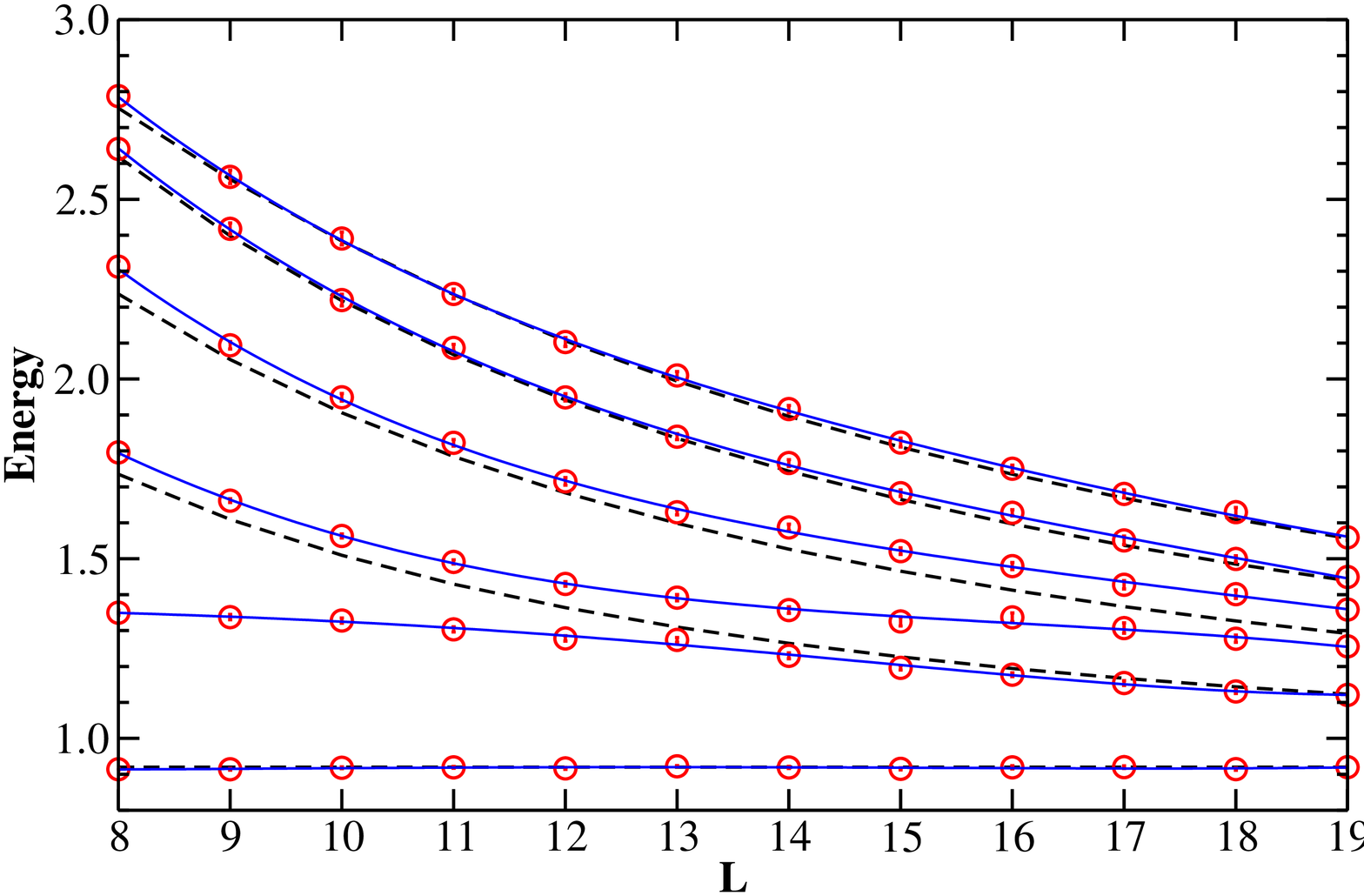}
\caption{(Left) Spectrum of the theory for different values of the
volume for the following simulation parameters:
$\nu=1.0$, $\lambda=1.4$, $m_\pi=0.36$.
The dashed lines describe the free two-particle spectrum.
(Right) The interpolated data using a polynomial.}
\label{L12spectr}
\end{figure}

In order to test the applicability of the two methods 
for different widths of resonance, we consider three different sets of 
parameters.
The first set is characterised by $\nu=1.0$, $\lambda=1.4$, $m_\pi=0.36$.
We tuned these parameters to have the intersection between the $\sigma$ energy
level and $(1,0,0)$ two-particle energy level around $L=12$. The physical mass
for the pion turns out to be $m^{ph}_\pi=0.460(2)$.
The spectrum (the first 6 levels) was
determined for different volumes ($8 \leq L \leq 19$) and the relative error
works out to be in the range 0.5\% - 1.0\% (see Figure~\ref{L12spectr} (Left)).

First of all, we interpolate the data for each level using three different
polynomials of order 3, 4 and 5 in order to have a relation between the
energy $E$ and the side box $L$ in the entire interval $[8,19]$ and
to provide a way to evaluate the systematic errors in our final results;
in Figure~\ref{L12spectr} (Right) we show one of these interpolations.
In Figures~\ref{L12spectr}, the dashed lines describing the free
two-particle spectrum are calculated using Eq.~\ref{eqElatt}.
Here we note that the mass $m_\pi$ in Eq.~\ref{eqElatt} is not the 
\emph{bare} mass.
The value measured on the lattice $m^{ph}_\pi=0.460(2)$ is taken as the 
\emph{physical} value; the relation between them is well-approximated by 
$m_\pi^{ph}=2 \sinh^{-1}( m_\pi /2)$.
\FIGURE{
  \includegraphics[width=0.36\textwidth,angle=-90]{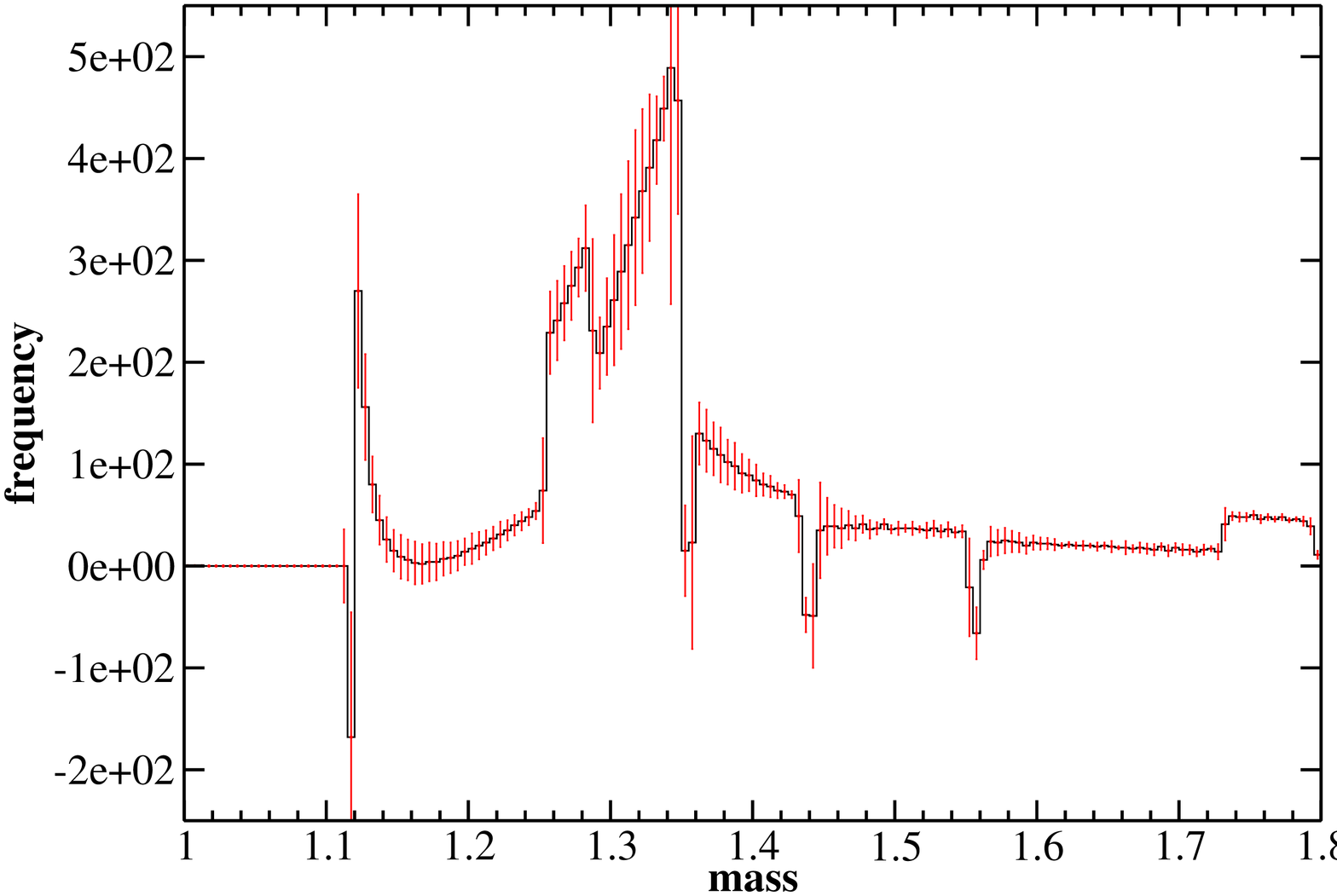}
\caption{The probability distribution $\tilde{W}=W-W_0$ obtained by data from
Figure~\protect\ref{L12spectr}.}
\label{L12histdiff}
}
 This free spectrum is used to determine
the distribution $W_0$ which is then subtracted to $W$ that is obtained from the
interacting spectrum.
It is important to note that if the number of levels considered to plot $W$
in the interacting spectrum are $N$, then the number of levels we have to
consider in the free spectrum to determine $W_0$ are just $N-1$.

Using the three previous polynomials we are able to produce a large number
of data (we fix $\Delta{L}=0.001$) that we can then use to get the probability
distribution $W$ described in Sec~\ref{subsec_propmeth} with the
corresponding systematic errors; fixing
the bin width to $\Delta{E}=0.005$ we obtain the histogram $\tilde{W}$ of
Figure~\ref{L12histdiff}. Note that to get $\tilde{W}$ both $W$ and $W_0$
are worked out from the same range with $L \in [8,19]$.
The error bars in Figure~\ref{L12histdiff} are the results of the systematic
errors coming from the histogram $W$ and the statistical errors coming from
the histogram $W_0$.

\begin{figure}[htb]
\hspace{-2mm}
  \includegraphics[width=0.36\textwidth,angle=-90]{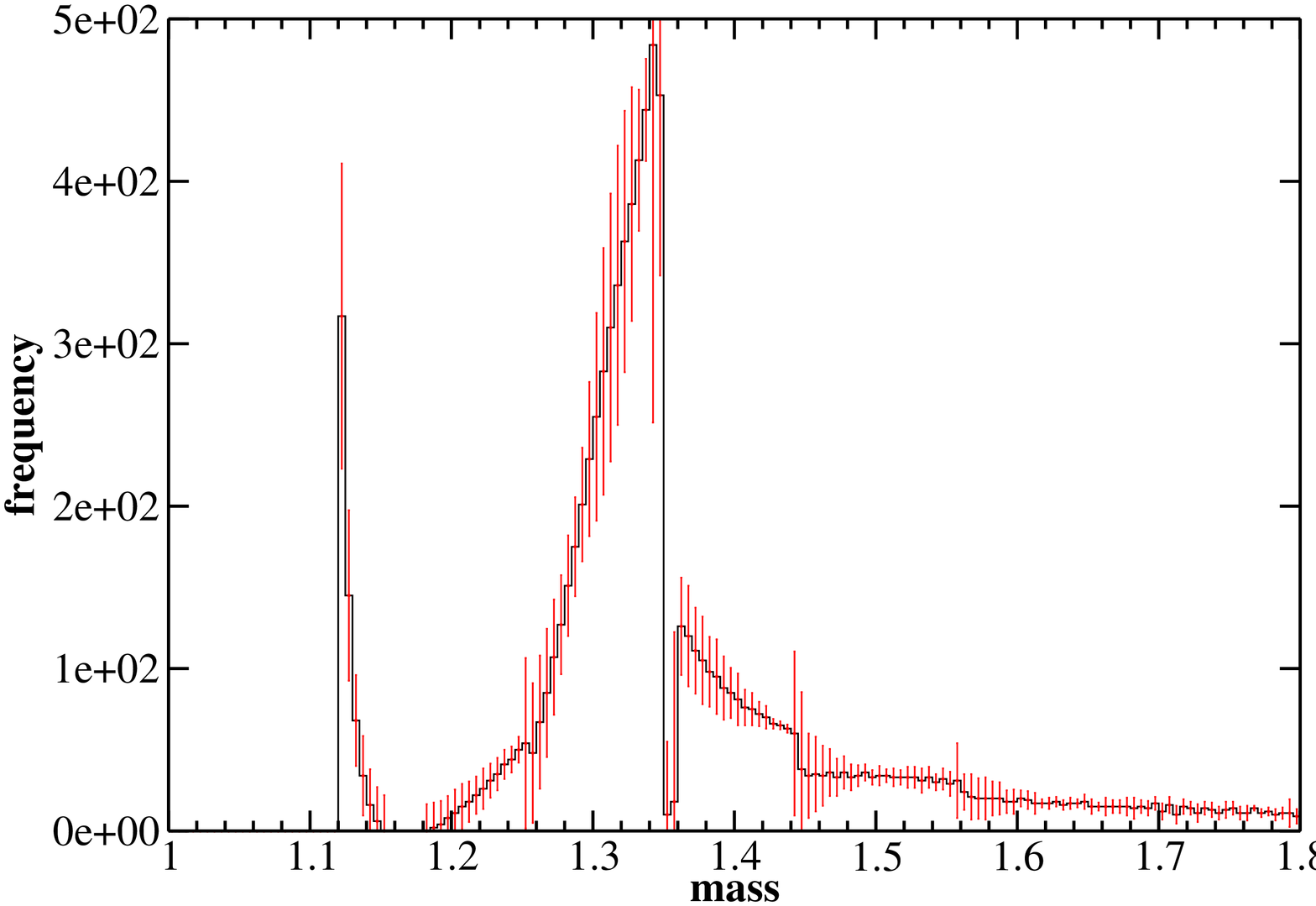}
\hspace{2mm}
  \includegraphics[width=0.36\textwidth,angle=-90]{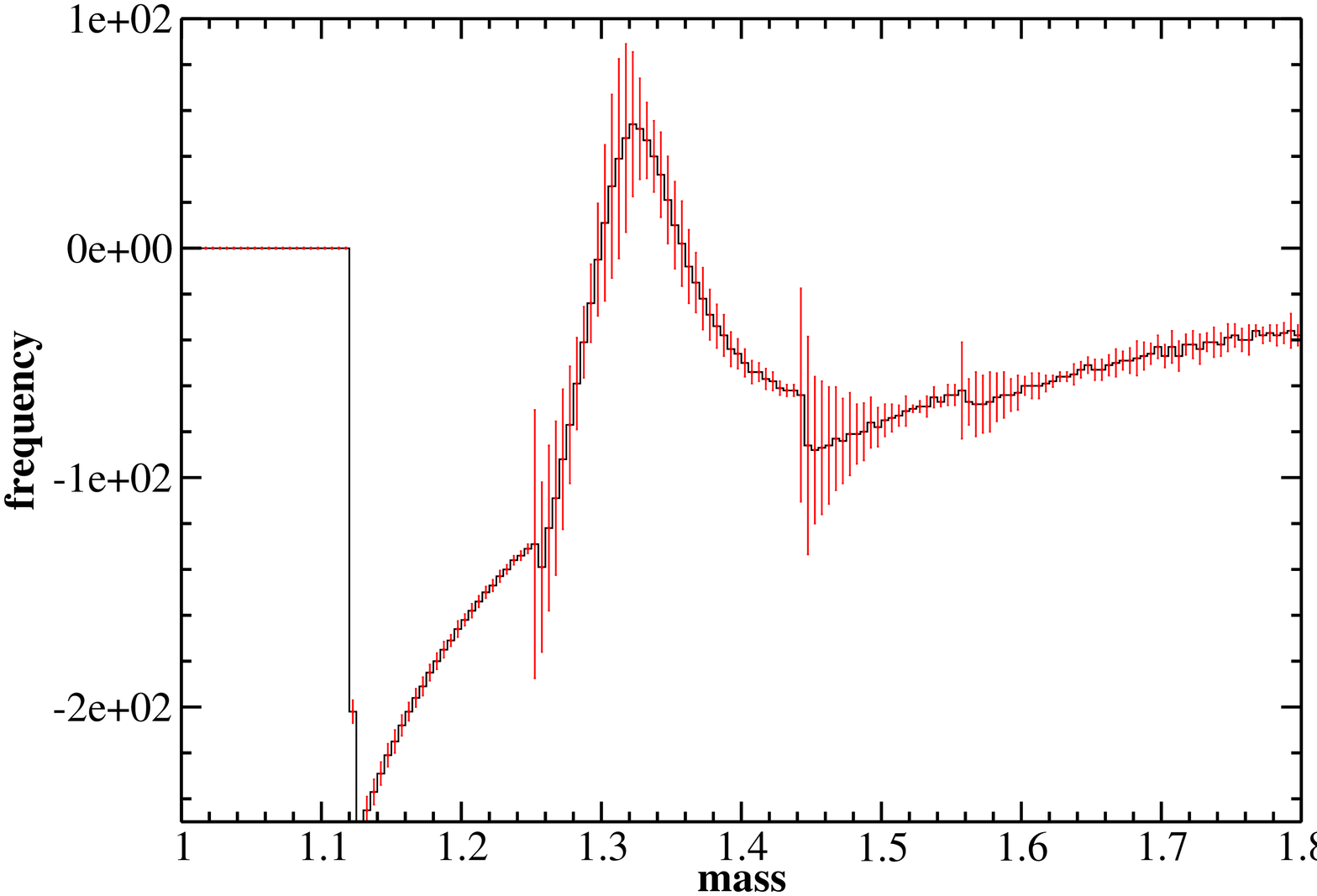}
\caption{(Left) Probability distribution $\tilde{W}$ obtained by data from
Figure~\protect\ref{L12spectrcorr}. (Right) Probability distribution $\tilde{W}$
obtained excluding from the analysis the two levels that in
Figure~\protect\ref{L12spectrcorr} are without a corresponding background.}
\label{L12histmod}
\end{figure}

Clearly, the shape of the histogram in Figure~\ref{L12histdiff} is far from 
the Breit-Wigner shape; the reason is related to the fact we are
considering only six energy levels but the conclusions of
Sec~\ref{subsec_propmeth} are true only in the limit of an infinite number of
levels. Moreover a lot on jumps and spikes are present. Our task is now to try
to improve this result in order to get more information from our raw data.
\FIGURE{
  \includegraphics[width=0.36\textwidth,angle=-90]{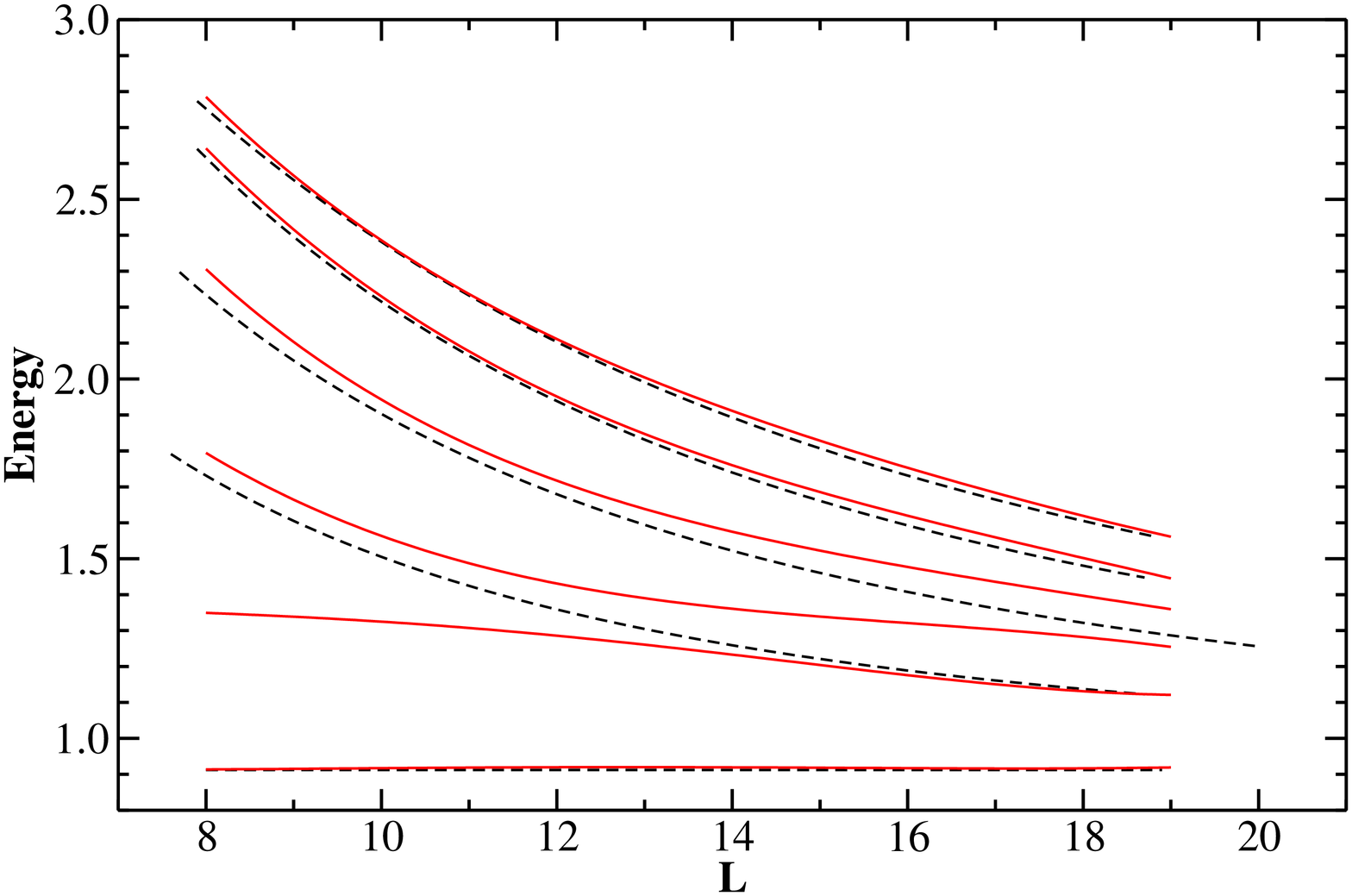}
\caption{Energy levels of Figure~\protect\ref{L12spectr} (Right) with the
correct free two-particle spectrum background.}
\label{L12spectrcorr}
}
We investigated the origin of the spikes and we understood they are
related to a ``wrong'' background $W_0$. It is easy to see
that the spikes appear every time there is the intersection between the
six levels of the interacting spectrum (or of the five levels of the free
spectrum) with the extremities of the volume range ($L=8$ and $L=19$).
Near those two extremities we have to be careful with what is the correct
background;
Figure~\ref{L12spectrcorr} shows a corrected background subtraction. 
In order to correctly subtract the free background, we lengthen each
free spectrum line. This is done so that the extremity of that line
has an energy equal to that of the extremity of the interacting spectrum
line closest to it. In this way all interacting lines are subtracte
correctly rather than the subtraction being affected by the limit
of the volume range that we are actually using in our simulations.
Using this procedure to determine $W_0$ we get the \emph{correct}
histogram of Figure~\ref{L12histmod} (Left).
Unfortunately, in Figure~\ref{L12histmod} (Left) we continue to see a jump
for $E \approx 1.35$; the origin of it can be understood looking at 
Figure~\ref{L12spectrcorr}. There are two extremity lines, one at $L=8$
and one at $L=19$ (both around $E \approx 1.35$), that are without
a ``background''; actually, in this case the background is the resonance
itself we are looking for.
Therefore, there is no way to avoid the presence of this jump because
we do not know anything about the resonance; the only thing we can do is
to completely exclude from our analysis those two levels, hoping that
the resonance can appear. In Figure~\ref{L12histmod} (Right) we show
the probability distribution $\tilde{W}$ in this last case; now clearly
a Breit-Wigner shape appears.

\begin{figure}[htb]
\hspace{-2mm}
  \includegraphics[width=0.36\textwidth,angle=-90]{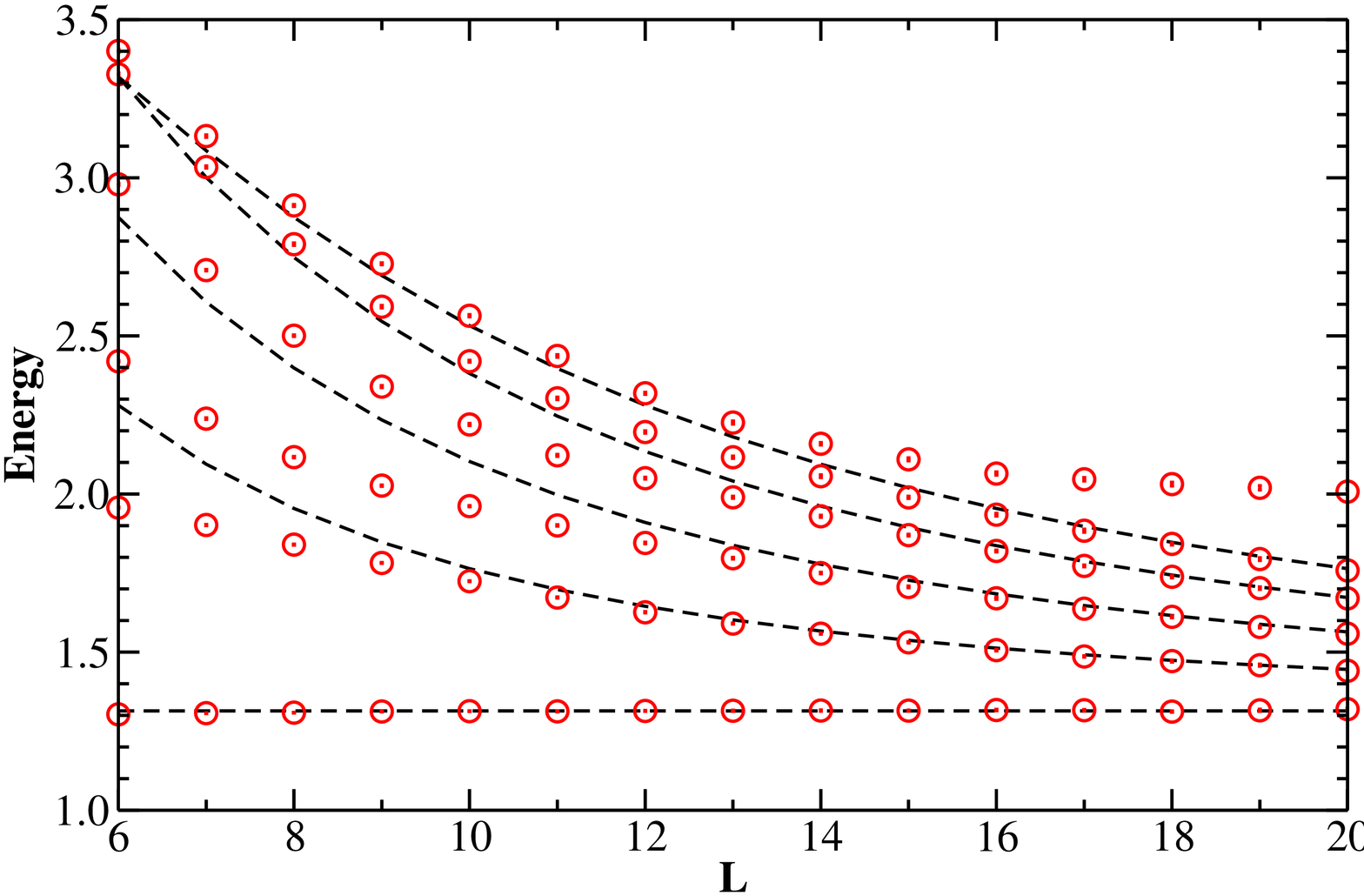}
\hspace{2mm}
  \includegraphics[width=0.36\textwidth,angle=-90]{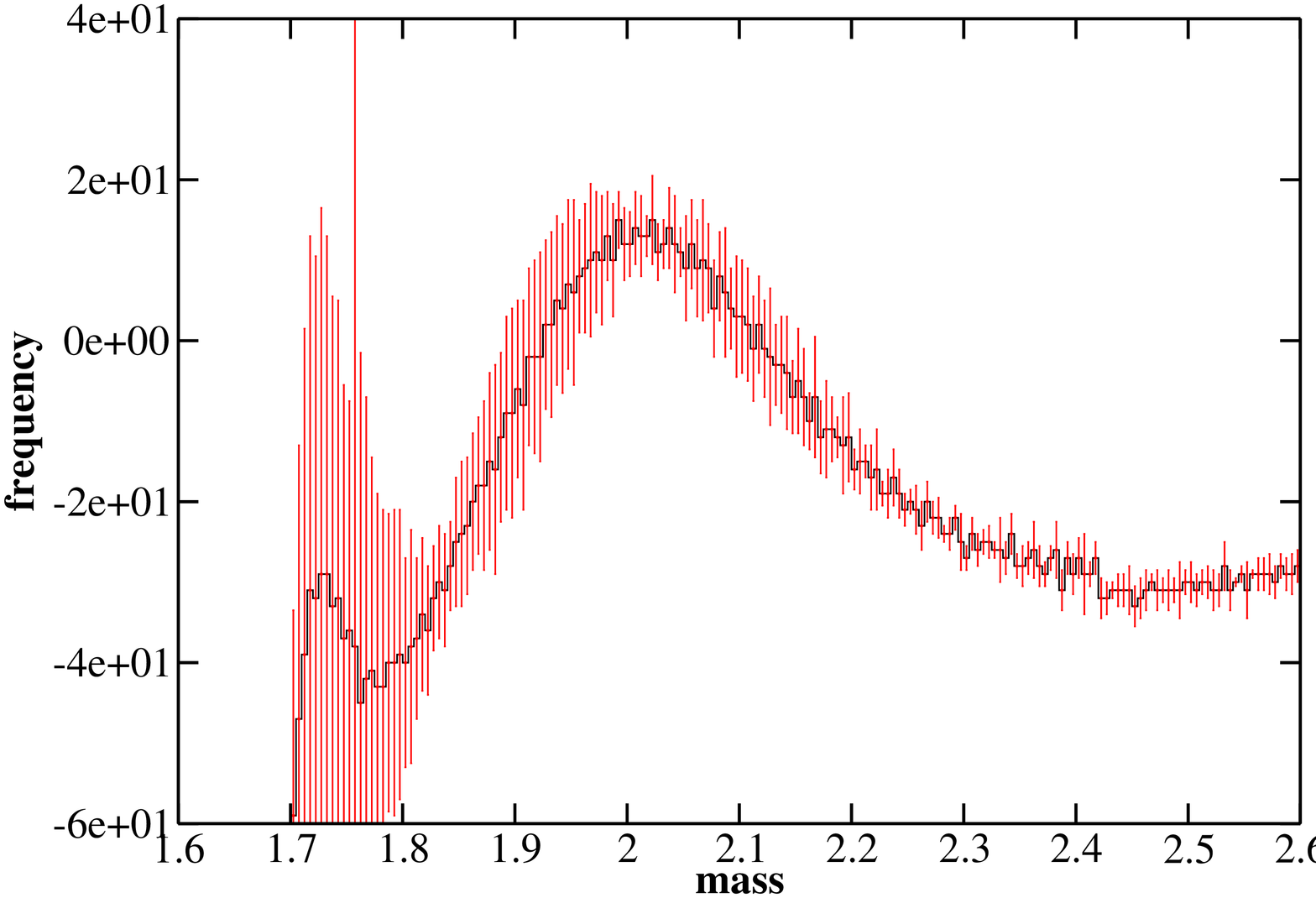}
\caption{(Left) Spectrum of the theory for different values of the
volume for the following simulation parameters:$\nu=1.0$, $\lambda=4.0$,
$m_\pi=0.56$. (Right) Probability distribution $\tilde{W}$ using the correct
background and excluding the two levels that are without a corresponding
background.}
\label{L8}
\end{figure}

It is now possible to fit these data to Eq.~\ref{bw} to determine the
parameters of the resonance; applying a sliding window procedure around
the peak, they turn out to be: $M_\sigma=1.330(5)$ and $\Gamma_\sigma=0.10(5)$.
We have simulated the theory with a second set of parameters, corresponding
to a larger width:
$\nu=1.0$, $\lambda=4.0$, $m_\pi=0.56$. In this case, we tuned them
to have the intersection between the $\sigma$ energy level and
$(1,0,0)$ two-particle energy level around $L=8$. The physical mass
for the pion turns out to be $m^{ph}_\pi=0.657(3)$.
In Figure~\ref{L8} (Left) we plot the spectrum for $6 \leq L \leq 20$
for the first
six levels; the relative error varies in the range 0.05\% - 0.2\%.
If we repeat all the procedure as described before (in particular we
exclude the two levels which are ``without'' background) we get the histogram
of Figure~\ref{L8} (Right); also in this case we can clearly see a
Breit-Wigner shape and we can fit these data obtaining the following
parameters: $M_\sigma=2.01(2)$, $\Gamma_\sigma=0.35(10)$.

\begin{figure}[htb]
\hspace{-2mm}
  \includegraphics[width=0.36\textwidth,angle=-90]{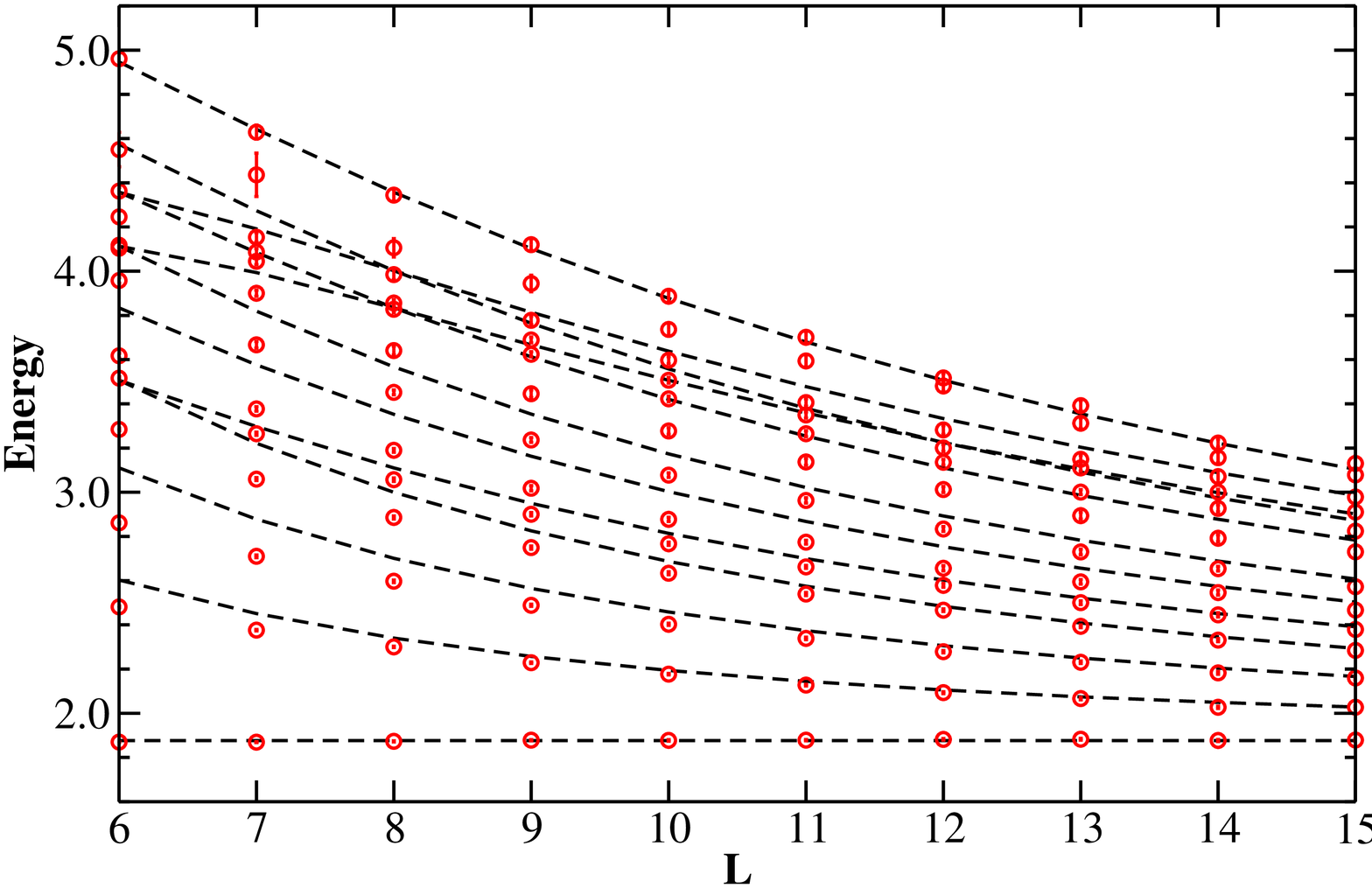}
\hspace{2mm}
  \includegraphics[width=0.36\textwidth,angle=-90]{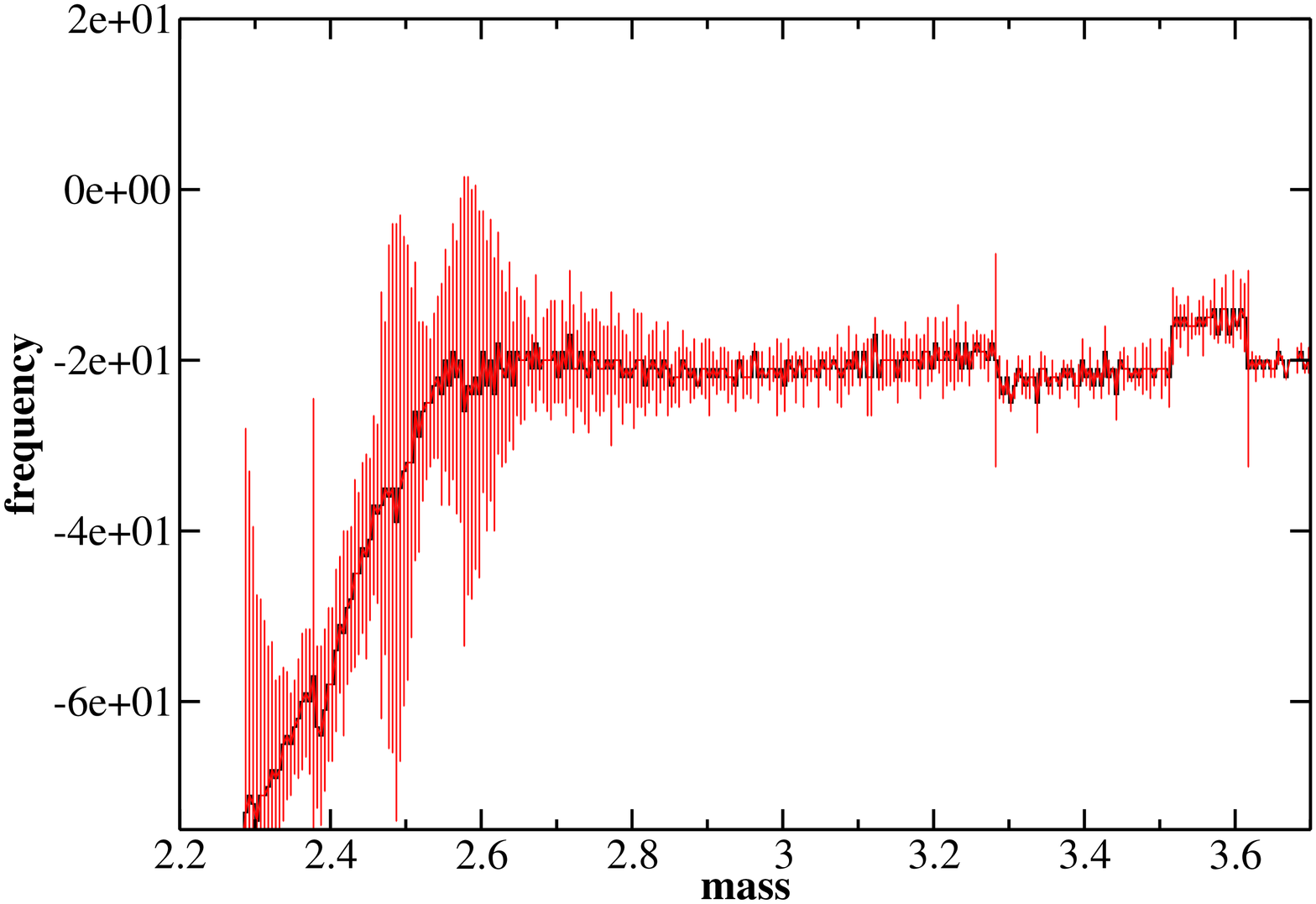}
\caption{Like Figure~\protect\ref{L8} but with simulation parameters:
 $\nu=1.0$, $\lambda=200.0$, $m_\pi=0.86$.}
\label{L10}
\end{figure}

Finally, we have run a third series of simulations with parameters
$\nu=1.0$, $\lambda=200.0$, $m_\pi=0.86$. They have been tuned to
have the intersection between the $\sigma$ energy level and
$(2,0,0)$ two-particle energy level around $L=10$. Because in this case
we are considering higher momentum, we expect the width of the resonance
is larger then the previous cases. In this case we take in account
13 levels to describe better the shape of the resonance.
In Figure~\ref{L10} (Left) the spectrum for $6 \leq L \leq 15$ is plotted;
the relative error varies in the range 0.15\% - 0.4\%.
The physical mass for the pion turns out to be $m^{ph}_\pi=0.938(3)$.
Unfortunately, as it is shown in Figure~\ref{L10} (Right) the probability
distribution plot is flat, i.e. no Breit-Wigner shape emerges.
It is clear that in this case, the only way to determine the parameters
of the resonance is to increase considerably the number of measurements
and consequently to decrease the relative errors in the spectrum determination.

In Table~\ref{table} a summary of our results for the three sets of parameters
are shown.

\begin{table}[h]
\begin{center}
\begin{tabular}{|l|l|l|l|l|}
\hline
Relative error in E(L) & $M_\sigma$  & $\delta(M_\sigma)/M_\sigma$  & $\Gamma_\sigma$ & $\delta(\Gamma_\sigma)/\Gamma_\sigma$ \\
\hline
0.5\%-1.0\% & 1.330(5) & 0.4\%   & 0.10(5)  &  50\%   \\
\hline
0.05\%-0.2\% & 2.01(2) & 1.0\%   & 0.35(10)  &  28\%   \\
\hline
0.15\%-0.4\% & -- & --   & --  &  --   \\
\hline
\end{tabular}
\end{center}
\caption{Results for the three sets of simulation parameters
with the corresponding relative errors.}
\label{table}
\end{table}
In applying L\"uscher's method to the data one only needs the original 
raw data; there is no need to fit it to a polynomial expression as in the 
histogram case. 
The first step is to convert the data on the energy spectrum to data on 
the momentum spectrum. This requires the dispersion relations. However 
should it be the lattice or continuum dispersion relations? The lattice 
dispersion relations are more natural, since they suppress lattice 
artifacts, but results were obtained for both below to emphasise how 
much more effective they are.
\begin{figure}[htb]
\hspace{-2mm}
  \includegraphics[width=0.36\textwidth,angle=-90]{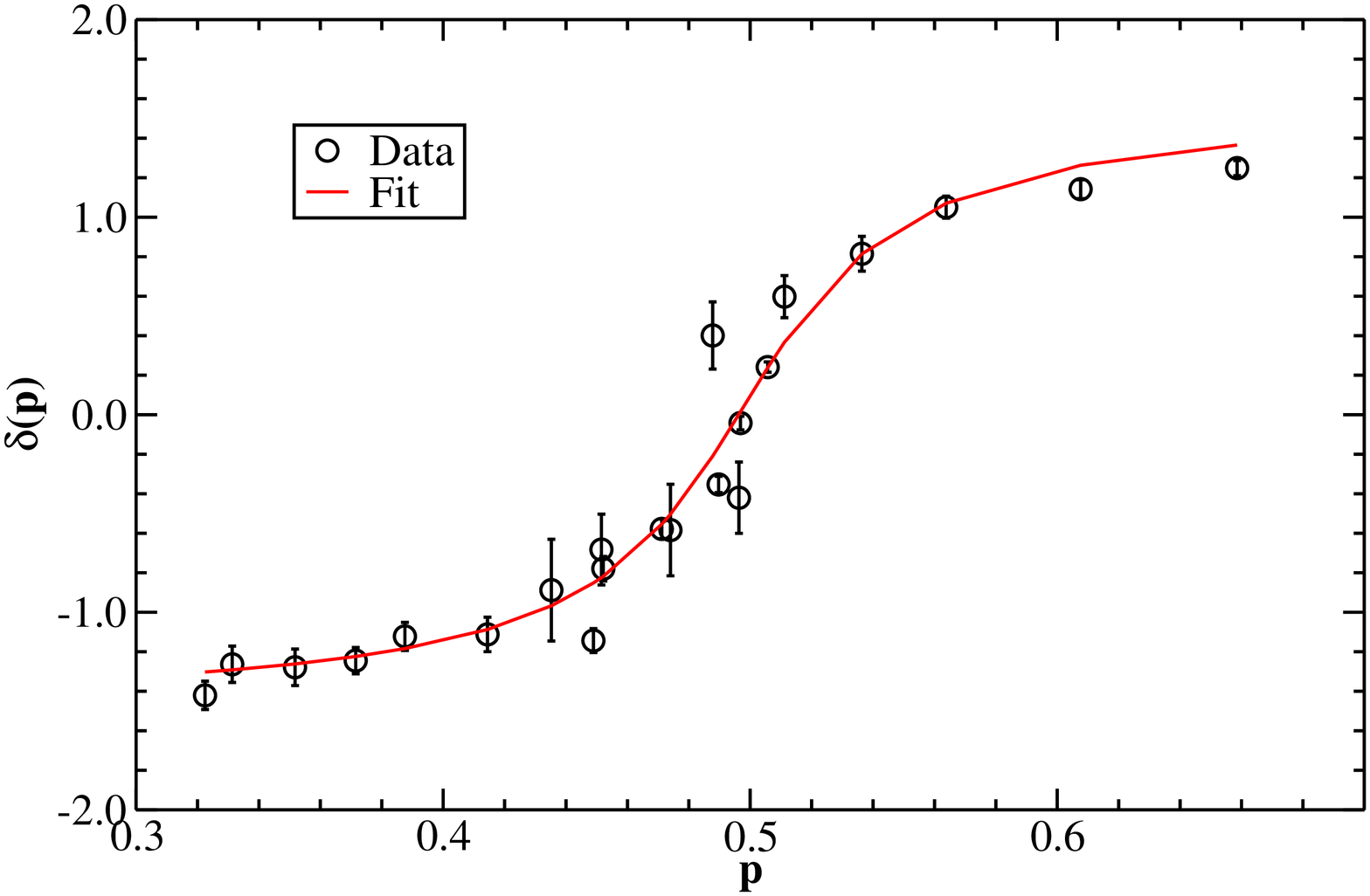}
\hspace{2mm}
  \includegraphics[width=0.36\textwidth,angle=-90]{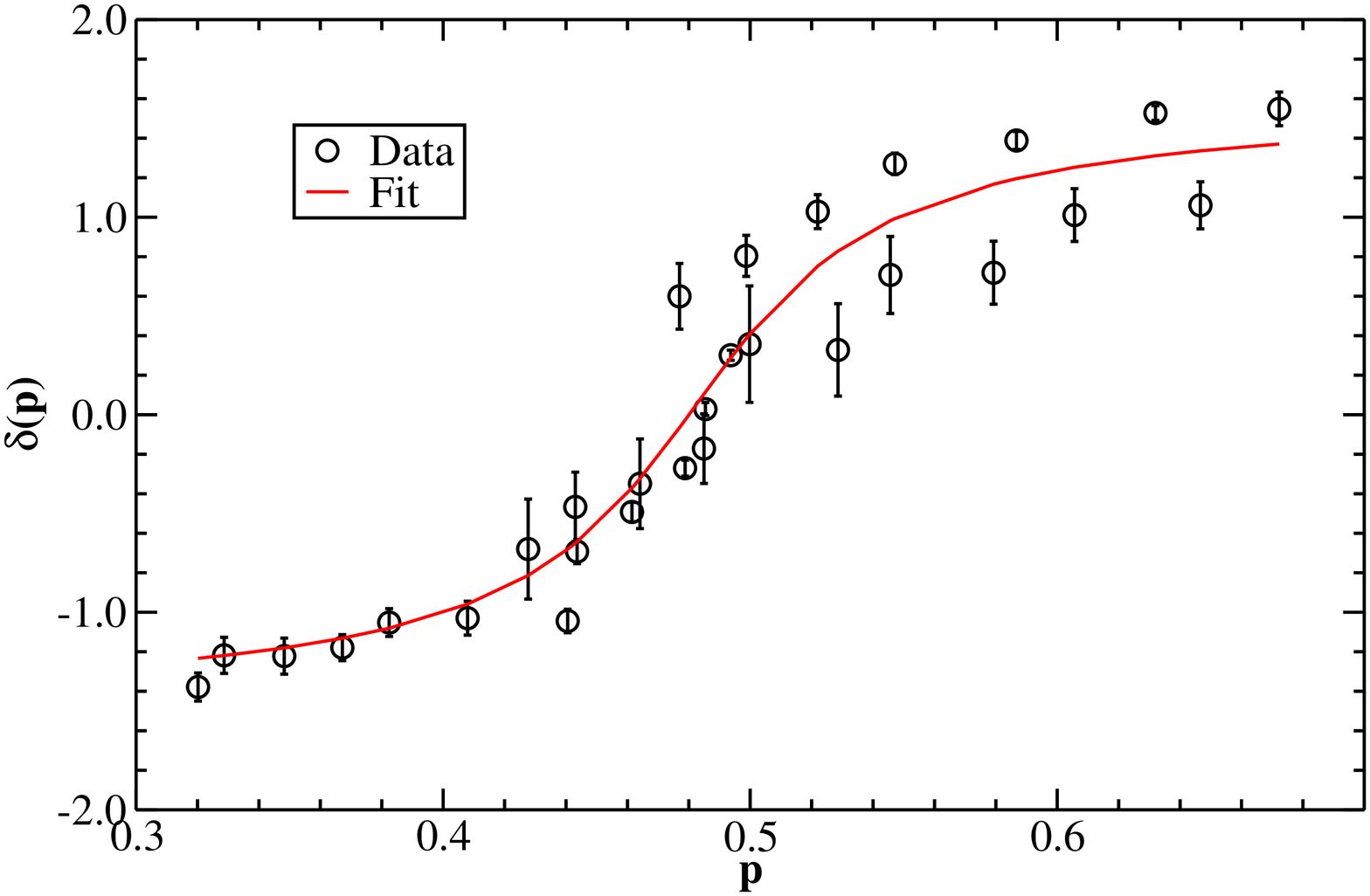}
\caption{(Left) $\delta(p)$ using Lattice dispersion relations at:
$\nu=1.0$, $\lambda=1.4$,
$m_\pi=0.36$. (Right) Same parameters, but with continuum dispersion 
relations. Both done with our approximation.}
\label{Dispersion}
\end{figure}
In order to use $p_n(L)$ and Eq.~\ref{lusherformu} to obtain $\delta(p)$, 
knowledge 
of $\phi(\kappa)$ is needed. The main difficulty here is the cumbersome 
definition of $\mathcal{Z}_{js}(1;q^{2})$. Not only that, but as 
mentioned the summation expansion given above does not even converge 
in the region required.
A more convenient definition is an integral representation of 
$\mathcal{Z}_{js}(1;q^{2})$ given in Appendix C of Ref~\cite{L\"uscher:1991a}. 
This expression can be used to numerically evaluate 
$\mathcal{Z}_{js}(1;q^{2})$. Some data on the values of $\phi(\kappa)$ 
can then be obtained from Eq.~\ref{tanlusherformu}. We fitted $\phi(\kappa)$
against these values to obtain an approximation of 
\begin{eqnarray}
\phi(\kappa) \approx (-0.09937)\kappa^{8} + (0.47809)\kappa^{6}  \\ \nonumber
+ (-0.62064)\kappa^{4} + (3.38974)\kappa^{2}
\label{eq:phi_fit}
\end{eqnarray}
The error between this approximation and $\phi(\kappa)$ is negligible 
compared with other errors.

From here we use Eq.~\ref{lusherformu} to obtain a profile of $\delta(p)$.
For the narrow case one can see the difference between use of the 
continuum dispersion relations and the lattice dispersion relations 
in Figure~\ref{Dispersion}.
\begin{table}[htb]
\begin{center}
\begin{tabular}{|l|l|l|}
\hline
\multicolumn{3}{|c|}{Results} \\
\hline
Parameters                   & $\phi(\kappa)$              & $\pi\kappa^{2}$ \\ \hline
$\nu = 1.0$, $\lambda = 1.4$ & $M_{\sigma} = 1.35(2)$       & $M_{\sigma} = 1.36(4)$\\
                             &$\Gamma_{\sigma} = 0.115(8)$  & $\Gamma_{\sigma} = 0.17(2)$ \\ \hline
$\nu = 1.0$, $\lambda = 4$ & $M_{\sigma} = 2.03(2)$ & $M_{\sigma} = 2.2(2)$\\
 & $\Gamma_{\sigma} =  0.35(2)$& $\Gamma_{\sigma} = 0.42(5)$ \\ \hline
$\nu = 1.0$, $\lambda = 200$ & $M_{\sigma} = 3.1(7)$ & $M_{\sigma} = 3(1)$ \\
 & $\Gamma_{\sigma} = 1.2(5)$ & $\Gamma_{\sigma} = 2(1)$ \\ \hline
\end{tabular}
\end{center}
\caption{Resonance mass and decay width using two different approximations
for $\phi(\kappa)$.}
\label{table2}
\end{table}
The lattice dispersion relations provide a tighter fit of the data, as 
well as having smaller errors. We also compared the use of the traditional 
approximation of $\phi(\kappa) = \pi\kappa^{2}$ with our approximation. 
After fitting, the results for the resonance mass and decay width in the 
two approximations are (both using lattice dispersion relations) shown in 
Table~\ref{table2}.

The errors are smaller when the approximation of Eq.~\ref{eq:phi_fit} are used, 
particularly for the broad resonance. It should also be noted that the two 
approximations effect the two resonance parameters differently. The dispersion 
relations have a more direct effect on $M_{\sigma}$ while the approximation of 
$\phi(\kappa)$ has a greater effect on $\Gamma_{\sigma}$. 
\begin{table}[htb]
\begin{center}
\begin{tabular}{|l|l|l|}
\hline
\multicolumn{3}{|c|}{Results} \\
\hline
Parameters & L\"{u}scher's Method & histogram method\\ \hline
$\nu = 1.0$, $\lambda = 1.4$ & $M_{\sigma} = 1.35(2)$ & $M_{\sigma} = 1.33(5) $\\
 & $\Gamma_{\sigma} = 0.115(8)$ & $\Gamma_{\sigma} = 0.10(5)$\\ \hline
$\nu = 1.0$, $\lambda = 4$ & $M_{\sigma} = 2.03(2)$ & $M_{\sigma} = 2.01(2)$\\
 & $\Gamma_{\sigma} = 0.35(2) $& $\Gamma_{\sigma} = 0.35(10)$ \\ \hline
$\nu = 1.0$, $\lambda = 200$ & $M_{\sigma} = 3.1(7)$ & $M_{\sigma} = N/A$ \\
 & $\Gamma_{\sigma} = 1.2(5)$ & $\Gamma_{\sigma} = N/A$ \\ \hline
\end{tabular}
\end{center}
\caption{A comparison between the L\"{u}scher and the histogram method.}
\label{table3}
\end{table}
This is because a 
cruder approximation of $\phi(\kappa)$ effects the profile of the scattering 
phase shift, which is related to the decay width. These results suggested 
it is optimal to use the lattice dispersion relations and our approximation.

\subsection{Comparison between the two methods}

The results for L\"uscher's method compared with the histogram method are 
shown in Table~\ref{table3}.
\FIGURE{
  \includegraphics[width=0.36\textwidth,angle=-90]{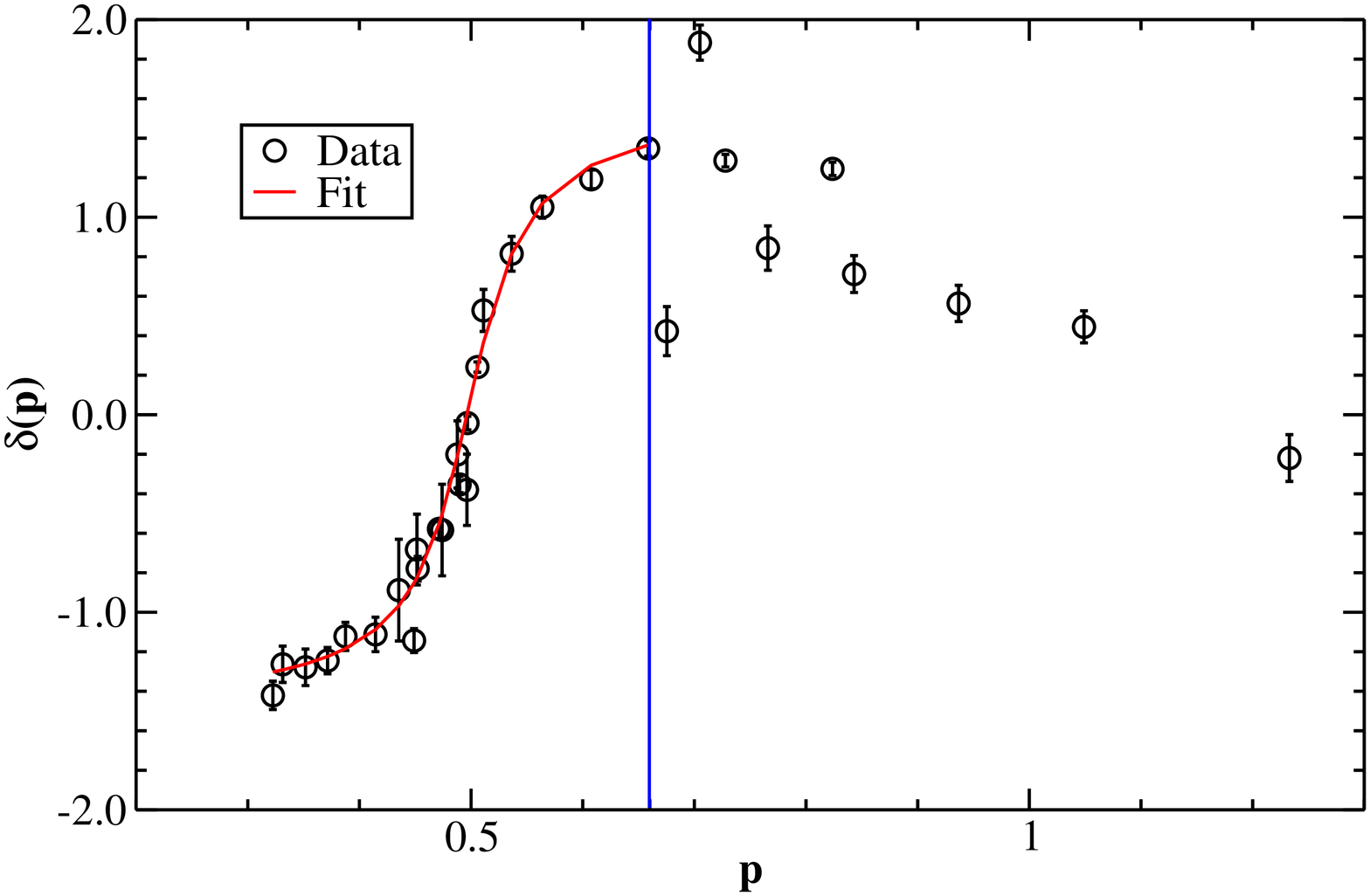}
\caption{Inelastic data with L\"uscher's formula. For the case of $\nu = 1.0$, $\lambda = 1.4$. (Onset of inelastic region marked).}
\label{Inelastic}
}
L\"uscher's method gives smaller errors than the histogram 
method, but the results are broadly consistent. L\"uscher's method 
manages to provide some estimate on the width of the resonance in the 
broad case. 

The broad resonance becomes a problem for the histogram 
method because there is no obvious peak to indicate the resonance mass 
(and hence no width of that peak to determine the decay width). 
One would need very precise data in order to avoid a washing out of
the structure of the histogram. 
L\"uscher's method also becomes more difficult to apply in the case 
of broad resonances. In the case of a broad resonance the profile 
of $\delta(p)$ is quite flat, hence a large range of parameters will 
be capable of fitting to the profile. Again an accurate determination 
of the energy levels is required to determine the profile precisely 
enough so that this is prevented.
Considering the amount of work necessary before one can use the 
histogram method (as detailed above), L\"uscher's method is considerably 
easier to apply, provided one has a good approximation of $\phi(\kappa)$. 
However, the histogram method can be used as a visual tool for 
spotting the resonance.

One restriction of L\"uscher's formula is that it only applies in the 
elastic region. An example of what happens in the inelastic region is 
provided in Figure~\ref{Inelastic}. It is possible that the histogram 
method will provide a means of determining the presence of a resonance 
in the inelastic region. Certainly a histogram can be constructed in the 
inelastic region, the only difficulty is that with the inapplicability 
of L\"uscher's formula it is unclear that the parameters of this 
histogram will have any relation to those of the resonance.

\section{Conclusions}

We have compared and contrasted the L\"uscher and the histogram methods. 
L\"uscher's method appears to both easier to apply and give 
smaller errors, however the histogram method does give results 
consistent with L\"uscher's method and does indeed visually indicate 
the presence of a resonance.

There are two major difficulties with both methods. First 
for broad resonances the relevant structure is washed 
out to some degree. For a histogram, the peak is hard to locate, while for the
fit, the profile of the phase shift is poorly constrained.
Secondly there is the inelastic region. L\"uscher's formula cannot be
used there. The histogram method can be applied to the data, but there is no 
argument that this is a sensible thing to do. There is also a difficulty in the 
general case, relevant to QCD, which has not been examined here. In the model 
above the resonance is clearly present in the channel, since this is an 
explicit feature of the Lagrangian of the model. In general however a 
resonance may not be so obvious and there is no reason a priori to expect that 
it will have a purely Breit-Wigner form.

\acknowledgments{This work is supported by Science Foundation Ireland under
  research grant 07/RFP/PHYF168. We are grateful for the continuing support of
    the Trinity Centre for High-Performance Computing, where the numerical
    simulations presented here were carried out.}

\end{document}